\documentclass[useAMS,usenatbib,fleqn]{mnras}

\usepackage{psfrag,amsmath,amssymb,color,epsfig,rotating}
\usepackage{graphicx}
\usepackage{graphics}
\usepackage{multirow}
\usepackage{tabularx}
\usepackage{geometry}
\usepackage{array}
\usepackage{comment}
\usepackage{setspace}% for double spacing
\usepackage[dvipsnames]{xcolor}

\usepackage[T1]{fontenc}
\usepackage{ae,aecompl}

\def\kk{\mathbf{k}}
\def\ka{\mathbf{k_1}}
\def\kb{\mathbf{k_2}}
\def\kc{\mathbf{k_3}}
\def\kta{\mathbf{\tilde{k}_1}}
\def\ktb{\mathbf{\tilde{k}_2}}
\def\ktc{\mathbf{\tilde{k}_3}}
\def\O{\mathcal{O}}
\def\R{\mathcal{R}}
\def\T{\mathcal{T}}
\def\P{\mathcal{P}}
\def\Rabc{\mathcal{R}_{\alpha\beta\gamma}}
\def\zh{\mathbf{\hat{z}}}
\def\xh{\mathbf{\hat{x}}}
\def\ph{\mathbf{\hat{p}}}
\def\kah{\mathbf{\hat{k}_1}}
\def\kahp{\mathbf{\hat{k}_{1 \perp}}}
\def\RB{\bar{R}}
\def\nn{\nonumber}
\def\eqn{Equation}

\title[]{Quantifying the Redshift Space Distortion of the Bispectrum I: Primordial Non-Gaussianity}

\author[S. Bharadwaj]{Somnath Bharadwaj$^{1,2}$\thanks{\href{mailto:somnath@phy.iitkgp.ernet.in}{somnath@phy.iitkgp.ernet.in}},  Arindam Mazumdar$^{1}$\thanks{\href{mailto:arindam.mazumdar@iitkgp.ac.in}{arindam.mazumdar@iitkgp.ac.in}}, Debanjan Sarkar$^{1}$\thanks{\href{mailto:debanjan@cts.iitkgp.ac.in}{debanjan@cts.iitkgp.ac.in}}
\\
$^{1}$Centre for Theoretical Studies, Indian Institute of Technology Kharagpur, Kharagpur - 721302, India\\
$^{2}$Department of Physics, Indian Institute of Technology Kharagpur, Kharagpur - 721302, India}

\date{}

\pubyear{2019}

\begin{document}
	\label{firstpage}
	\pagerange{\pageref{firstpage}--\pageref{lastpage}}
	\maketitle
	
	\begin{abstract}
		The anisotropy of the redshift space bispectrum contains a wealth of cosmological information. This anisotropy depends on the orientation of three vectors $\ka,\kb,\kc$ with respect to the line of sight. Here we have decomposed the redshift space bispectrum 
		in spherical harmonics which completely quantify this anisotropy. To illustrate this we consider linear redshift space distortion of the bispectrum arising from primordial non-Gaussianity. In the plane parallel approximation only the first four even $\ell$ multipoles have non-zero values, and we present explicit analytical expressions for all the non-zero multipoles {\it i.e.} upto $\ell=6,m=4$.  The ratio of the different  multipole moments to the real space bispectrum depends only on $\beta_1$ the linear redshift distortion parameter and the shape of the triangle. Considering triangles of all possible shapes, we have studied how this ratio depends on the shape of the triangle for  $\beta_1=1$. We  have also studied the $\beta_1$ dependence for some of the extreme triangle shapes. If measured in future, these multipole moments  hold the potential of constraining $\beta_1$.  The results presented here are also important if one wishes to  constrain  $f_{\text{NL}}$ using redshift surveys. 
	\end{abstract}
	
	
	%%%%%%%%%%%%%%%%%%%%%%%%%%%%%%%%%%%%%%%%%%%%%%%%%%
	
	%%%%%%%%%%%%%%%%% BODY OF PAPER %%%%%%%%%%%%%%%%%%
	
	%***********************************************************************************
	\section{Introduction}
	\label{sec:intro}
	Redshift space distortion (RSD) caused  by peculiar velocities introduces a very distinct pattern in the observations of  large-scales structures (LSS) in the Universe. This is present in all  the LSS tracers where  distances  are inferred from  redshifts, e.g.  galaxies, quasars, the Lyman-$\alpha$ forest and the cosmological 21-cm signal. At large length scales coherent inflows into over-dense regions cause these to appear enhanced and squashed along the line of sight (LoS) direction whereas outflows from under-dense regions cause these to appear more under dense  and elongated along the LoS (Kaiser effect; \citealt{kaiser87}). At small length-scales, random motions cause the structures to appear elongated along the LoS (Finger of God effect; \citealt{jackson72-FoG}). The net effect is that the clustering pattern, which is expected to be statistically isotropic in real space (where the actual distances are known),  appears anisotropic relative to the LoS direction in redshift space (see \citealt{hamilton-98-rsd-review} for a review).

	The anisotropy of the redshift space two-point correlation function, or equivalently the power spectrum, is well studied in the literature. Considering a biased tracer, 
	the redshift space power spectrum $P^s(\kk_1)$ in linear theory (valid on large scales) can be expressed as an enhancement of the isotropic real space power spectrum $P^r(k_1)$ by a factor  $(1+\beta_1 \mu_1^2)^2$ which is known as the Kaiser enhancement factor \citep{kaiser87}, where $\beta_1=f/b_1$ is the linear redshift distortion parameter which is the ratio of $f(\Omega_m)$ the logarithmic derivative of the  growth rate of linear density perturbations  and the linear bias $b_1$, and $\mu_1=\zh \cdot \ka /k_1$ is the cosine of the angle between the wavevector $\ka$ and LoS direction $\zh$.
	Measurements of $P^s(\kk_1)$ at large scales can be used to constrain $f$ as a function of redshift \citep{loveday96, peacock01, hawkins03, guzzo-pierleoni08}. In addition to $f$, $P^s(\kk_1)$ at relatively small (weakly non-linear) scales contain a number of important cosmological information, e.g. the signature of massive neutrinos \citep{hu98-neutrino}, which requires a careful modelling of $P^s(\kk_1)$. 
	Several models based on higher order perturbation theory \citep{heavens98-NL-RSD, scoccimarro04,matsubara08,Taruya10-TNS} and effective field theory \citep{desjacques18-RSD-EFT} have been proposed to describe $P^s(\kk_1)$ at weakly non-linear scales.
	At small (non-linear) scales, where the Finger of God (FoG) effect is important, a number of models have been considered where the FoG suppression is modelled by considering a damping term along with the Kaiser enhancement \citep{davis-peebles83, peacock92,park-vogeley94, ballinger-peacock-heavens96, hatton-cole99, seljak01-RSD-halo-model, white01-RSD-halo-model,bharadwaj01-nonlinear-RSD}. It is convenient to study the anisotropic $P^s(\kk_1) \equiv P^s(k_1,\mu_1)$ in terms of angular multipoles $P^s_{\ell}$ which define the decomposition of $P^s(k_1,\mu_1)$ into Legendre polynomials $\P_{\ell}(\mu)$ as $P^s(k_1,\mu_1)=\sum_{\ell} P^s_{\ell}(k_1) \P_{\ell}(\mu_1)$ \citep{hamilton-92-RSDcorrelation,cole-fisher-weinberg94}.
	In linear theory under the plane parallel approximation, only the first three even moments namely monopole ($\ell=0$), quadrupole ($\ell=2$) and hexadecapole ($\ell=4$) are non-zero  \citep{cole-fisher-weinberg94}. Also, in linear theory, the ratio $P^s_{2}/P^s_{0}$ reduces to 
	$(4\beta_1/3+4 \beta_1^2/7)(1+2\beta_1/3+\beta_1^2/5)^{-1}$ which gives a direct measure of $\beta_1$ \citep{cole-fisher-weinberg94}. Since RSD is a direct consequence of the velocity field, it is  sensitive to the density potential fluctuation and thus can be used to test  theories  of modified gravity \citep{linder08-RSD-GR, song09-RSD, de_la_torre16-gravity-test-from-RSD-and-lensing, johnson-blake-16-modified-gravity-using-galaxy-peculiar-velocities, mueller18-BOSS-RSD}.

	The power spectrum is adequate to fully characterises the statistical properties  of the clustering pattern 
	if it is a Gaussian random field. The simplest models of inflation  predict the primordial fluctuations to be a Gaussian random field \citep{baumann09-inflation}, the density fluctuations are however predicted to become non-Gaussian as they evolve (induced non-Gaussianity; \citealt{fry84-bispec}) 
	due to the non-linear growth and non-linear biasing.  Further,  several inflationary scenarios predict 
	the primordial fluctuations to be non-Gaussian (primordial non-Gaussianity; \citealt{bartolo04-PNG-rev}). It  is then 
	necessary to consider higher order statistics,  the three-point correlation function or its Fourier conjugate the bispectrum being  the lowest order statistic sensitive to non-Gaussianity. 
	Measurements of bispectrum from the observations of Cosmic Microwave Background (CMB) \citep{fergusson12-PNG,oppizzi18-PNG, planck18png, shiraishi19-PNG} and galaxy surveys \citep{feldman01-bispec, scoccimarro04-PNG, ligouri10-PNG, ballardini19png} have been used to place tight constraints on primordial non-Gaussianity. Second order perturbation theory predicts \citep{matarrese97} that measurements of the bispectrum in the weakly non-linear regime can be used to determine the bias parameters, and this has been employed in the galaxy surveys to quantify the galaxy bias parameters \citep{feldman01-bispec, scoccimarro01-IRAS-bispec, verde02-2DF, nishimichi07, gilmartin15-bispec}. Further, the measurements of bispectrum enable us to lift the degeneracy between $\Omega_m$ (which appears in $f(\Omega_m)$) 
	and $b_1$, something  which is not possible by considering only the power spectrum \citep{scoccimarro99-RS-bispec}.

	Like the power spectrum, the anisotropy of the redshift space bispectrum contains a wealth of cosmological informations. It is therefore quite important to accurately model and quantify this.   
	\citet{Hivon1995} and  \citet{Verde1998} have calculated the bispectrum in redshift space. However,  the focus in these works has been on  measuring the large scale bias and the cosmological parameters, and they have not quantified the anisotropy arising from redshift space distortion. 
	\citet{scoccimarro99-RS-bispec} have quantified the  redshift space anisotropy of the bispectrum by decomposing it into spherical harmonics.  However, beyond the monopole the analysis is restricted to only one of the quadrupole components $(\ell=2,m=0)$. \citet{Hashimoto2017} also have considered the  single quadrupole component of the redshift space bispectrum. The works mentioned above all  use non-linear perturbation theory to calculate the  induced redshift space bispectrum   arising from Gaussian initial perturbation.  The results are also extensively validated using large N-body simulations. However  the anisotropy arising from redshift space distortions has only been partly analysed, the analysis being restricted to a single quadrupole component and a very restricted set of triangle configurations. 
	In a more recent work, \citet{Slepian2018} present a technique to quantify the redshift space three-point correlation function by expanding it in terms of products of two spherical harmonics. 
	In a very recent work \citet{Sugiyama2019a} have used a tri-polar spherical harmonic decomposition to quantify the anisotropy of the redshift space bispectrum, and they demonstrate this technique by applying it to the Baryon Oscillation Spectroscopic Survey (BOSS) Data Release 12. The last two works  mentioned here present very efficient computational techniques which are well suited for large galaxy surveys.

	In the present  work we focus on quantifying the anisotropy of the redshift space   bispectrum. 
	The bispectrum in real space (as against redshift space) only depends on the shape and size  of the triangle 
	$\ka,\kb,\kc$ and is independent of how the triangle is oriented. The redshift space bispectrum, however, also depends on the orientation of the triangle through  $\mu_1,\mu_2,\mu_3$ where $\mu_a=\zh \cdot \mathbf{k}_a/k_a$ with $a=1,2,3$. 
	The issue here is to fix the shape and size of the triangle,  and 
	quantify the joint $\mu_1,\mu_2,\mu_3$ dependence of the redshift space bispectrum.  We show that the redshift space bispectrum can be expressed as $B^s(k_1,\mu,t,\ph)$ where the parameters $k_1$ and $\mu,t$ (defined later) respectively quantify the size and the shape of the triangle, while the orientation of the three vectors $\ka,\kb,\kc$ with respect to $\zh$ is quantified through 
	the unit vector $\ph$ which  has components $p_z=\zh \cdot \kah$ and $p_x=\zh \cdot \kahp$. Here $\kah$ 
	is an unit  vector along $\ka$ which is taken to be the largest side of the triangle, 
	and $\kahp$ is an unit vector perpendicular to $\ka$ in the plane of the triangle. 
	We have quantified the anisotropy of the redshift space bispectrum by decomposing it in spherical harmonics $Y^m_{\ell}(\ph)$. The multipole moments $\bar{B}^m_{\ell}(k_1,\mu,t)$ provide a relatively simple and straight forward method to quantify the redshift space distortion of the bispectrum. In order to illustrate this method we have considered the bispectrum from primordial non-Gaussianity where the linear theory of redshift space distortion can be applied. In this case $\RB^m_{\ell}(\beta_1,\mu,t)$ which is the ratio of multipole moment  $\bar{B}^m_{\ell}(k_1,\mu,t)$ to the real space bispectrum $B^r(k_1,\mu,t)$ is independent of $k_1$ the size of the triangle, and it depends  only on $\beta_1$  the linear redshift distortion parameter and $\mu,t$ the shape of the triangle. Here we provide analytical expressions for all the non-zero $\RB^m_{\ell}(\beta_1,\mu,t)$ 
	{\it i.e.} up to $(\ell=6,m=4)$. We study the shape dependence of $\RB(\beta_1,\mu,t)$  considering triangles of all possible shapes for $\beta_1=1$. We have also studied the $\beta_1$ dependence for a few extreme triangle shapes and briefly discuss the possibility of using  observations to measure $\beta_1$. 
	We plan to carry out a similar analysis for the induced bispectrum arising from non-linear evolution and 
	present the results in a subsequent paper. 
	A brief outline of the present paper follows. 
	
	Section~\ref{sec:triangle-config} presents the parameters that we use to quantify  the shape, size and orientation  of a triangle, while our method to quantify the anisotropy of the redshift space bispectrum is presented in  Section~\ref{sec:RSD}. Our results for linear redshift space distortions are presented in Section~\ref{sec:linear-RSD}, and these are discussed in Section~\ref{sec:discussion}. Results for some of the multipole moments which are not included  in Section~\ref{sec:linear-RSD} have been presented in an Appendix.   
	
	%***********************************************************************************
	\section{Parameterizing  Triangle Configurations}
	\label{sec:triangle-config}
	
	The bispectrum is defined as
	%-------------------
	\begin{equation}
	B(\ka,\kb,\kc)=V^{-1}\,  \langle \Delta(\ka) \Delta(\kb) \Delta(\kc) \rangle 
	\,,
	\label{eq:a1}
	\end{equation}
	%-------------------
	with the condition $\ka+\kb+\kc=0$ which ensures that the three vectors form a closed triangle. 
	Further,  $B^r(\ka,\kb,\kc)$ the real space bispectrum (as against redshift space) 
	is independent of how the triangle is oriented in space, and it depends only on the shape and size of the triangle. In order to uniquely quantify  the shape and size of any triangle we label the sides such that 
	%-------------------
	\begin{equation}
	k_1 \ge k_2 \ge k_3     
	\label{eq:a2}
	\end{equation}
	%-------------------
	where $k_i= \mid \mathbf{k}_i \mid$.

\begin{figure}
		\psfrag{X}[c][c][1.2][0]{$X$}
		\psfrag{Y}[c][c][1.2][0]{$Y$}
		\psfrag{Z}[c][c][1.2][0]{$Z$}
		\psfrag{k1}[c][c][1.3][0]{$\mathbf{\tilde{k}_1}$}
		\psfrag{k2}[c][c][1.3][0]{$\mathbf{\tilde{k}_2}$}
		\psfrag{k3}[c][c][1.3][0]{$\mathbf{\tilde{k}_3}$}
		\psfrag{theta}[c][c][1.3][0]{$\theta$}
		\psfrag{chi}[c][c][1.2][0]{$\chi$}
		
		\centering
		\includegraphics[width=0.3\textwidth,angle=-90,trim=2cm 0.5cm 1.5cm .5cm,clip]{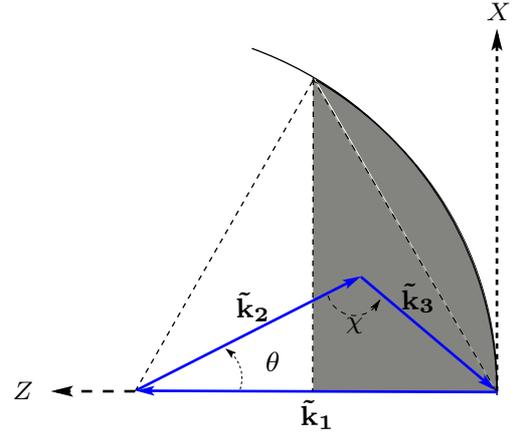}
		\caption{This shows the allowed configurations for the reference triangle.}
		\label{fig:triangle_config}
	\end{figure}

\begin{figure}
		%\psfrag{mu}[c][c][1][0]{$\mu$}
		%psfrag{t}[c][c][1][0]{$t$}
		\psfrag{Squeezed}[2][c][1.1][0]{Squeezed}
		\psfrag{cos chi}[c][c][1.1][0]{$\cos \chi$}
		\psfrag{Equilateral}[c][c][1.1][0]{Equilateral}
		\psfrag{S Isosceles}[c][c][1.1][0]{S Isosceles}
		\psfrag{L Isosceles}[c][c][1.1][0]{L Isosceles}
		\psfrag{Linear}[c][c][1.2][0]{Linear}
		\psfrag{Stretched}[c][c][1.1][0]{Stretched}
		
		\centering
		\includegraphics[width=0.48\textwidth,angle=-90,trim=1cm 6cm 1.5cm 1cm,clip]{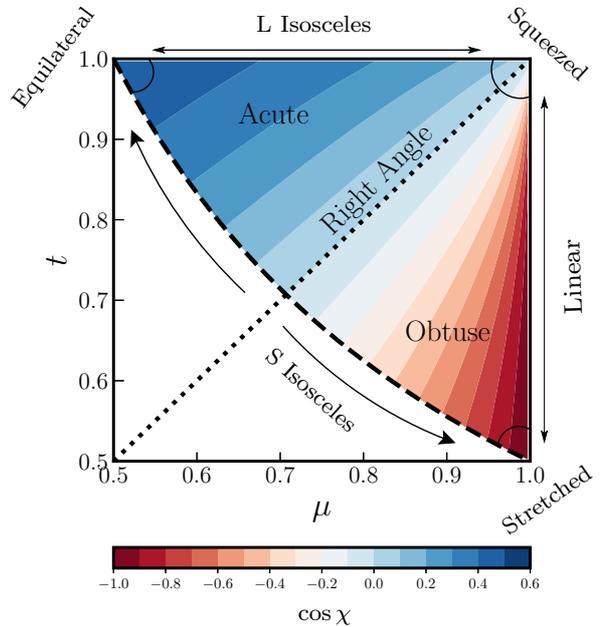}
		\caption{This shows the allowed $\mu,t$ space along with the triangle shapes corresponding to different values of $\mu$ and $t$. The contours correspond to different values of $\cos \chi$.}
		\label{fig:allowed_regn}
	\end{figure}

	We next identify a reference triangle $\kta,\ktb,\ktc$ in the $x-z$ plane (Figure~\ref{fig:triangle_config}) which is related to $\ka,\kb,\kc$ through a rigid body rotation  which maps 
	$(\kta,\ktb,\ktc)$ to  $(\ka, \kb, \kc)$ where 
	%-------------------
	\begin{equation}
	\kta=k_1 \, \zh \,.    
	\label{eq:a3}
	\end{equation}
	%-------------------
	We use the length of the largest side $k_1$ to specify the size of the triangle. Considering $\ktb$, 
	we express  this as 
	%-------------------
	\begin{equation}
	\ktb=t k_1 [-\mu  \, \zh + \sqrt{1-\mu^2} \, \xh]
	\label{eq:a4}
	\end{equation}
	%-------------------
	where $t=k_2/k_1$ specifies $k_2$ in units of $k_1$ 
	and $\mu=\cos \theta =-\kta\cdot \ktb/(k_1 k_2)$ is the cosine of the angle between $-\kta$ and $\ktb$. We use the parameters $\mu$ and $t$ to specify the shape of the triangle. 
	From \eqn~(\ref{eq:a2}) it follows that $\ktb$  is restricted to lie within the shaded region shown in Figure~\ref{fig:triangle_config}. This implies that the values of $\mu$ and $t$ are restricted to the range 
	%-------------------
	\begin{equation}
	0.5 \le t, \mu  \le 1  \, {\rm with} \, t \mu \ge 0.5  \,.
	\label{eq:a5}
	\end{equation}
	%-------------------
	Triangles of all possible shapes are uniquely represented by points in the allowed region of the $\mu,t$ configuration space shown in  Figure~\ref{fig:allowed_regn}. We use the parameters $k_1,\mu,t$ to represent the shape and size of all possible triangles, and express the bispectrum as $B^r(k_1,\mu,t)$.

	%------------------------------------------------------------------
	%------------------------------------------------------------------
	
	We now briefly discuss the shapes corresponding to the different $\mu,t$ values shown in Figure~\ref{fig:allowed_regn}. The right boundary $\mu=1$ corresponds to linear triangles 
	($\chi=180^{\circ}$ in Figure~\ref{fig:triangle_config})
	where $\kta,\ktb,\ktc$ are all aligned.
	The limit $\mu \rightarrow 1$ and  $t \rightarrow 0.5$ converges to  stretched triangle where $\ktb=\ktc=-\kta/2$, and the limit $\mu \rightarrow 1$ and  $t \rightarrow 1$ converges to squeezed  triangle where $\ktb=-\kta$ and   $\ktc=0$. The upper boundary $t=1$ corresponds to the L isosceles triangles  ($k_1=k_2$) where the two  larger sides are of equal length,  whereas the lower boundary $2 \mu t=1 $  corresponds to the S isosceles triangles $(k_2=k_3)$  where the two  smaller sides are of equal length.  
	The limit $\mu \rightarrow 0.5$ and  $t \rightarrow 1$ converges to  the equilateral triangle which separates the L and S isosceles triangles. Contours corresponding to various values of $\cos \chi$ are shown for reference. 
	The line $\mu=t$ corresponds to right angle triangles ($\cos \chi=0$),  whereas $\mu >t$ and $\mu <t$ respectively correspond to obtuse $(\cos \chi <0 $) and acute $(\cos \chi > 0$) triangles. 
	
	%-------------------
	%-------------------
	
	Starting from a reference triangle $(\kta, \ktb, \ktc)$ in the $x-z$ plane it is possible to obtain a triangle $(\ka, \kb, \kc)$ of the same shape and size but with a different  spatial orientation through a rigid body rotation  
	$\R_{\alpha\beta\gamma}:(\kta,\ktb,\ktc) \rightarrow{} (\ka, \kb, \kc)$ where  $\alpha,\beta,\gamma$  are three angles which parameterize all possible rigid body rotations. Here  we represent rigid body rotations using $\alpha$  which is the rotation angle  and  $\mathbf{\hat{n}}$ which is an  unit vector that denotes the rotation axis. 
	The angles $\beta$ and $\gamma$ denote the direction of  $\mathbf{\hat{n}}$ which has components $\hat{n}_x=\sin \beta  \, \cos \gamma$, $\hat{n}_y=\sin \beta  \, \sin \gamma$ and $\hat{n}_z=\cos \beta$.  
	The transformation $\mathbf{V}=\Rabc \mathbf{\tilde{V}}$ of an arbitrary vector  $\mathbf{\tilde{V}}$ under this rotation can be expressed as 
	%-------------------
	\begin{equation}
	\mathbf{V}=\cos \alpha \, \mathbf{\tilde{V}}    - \sin \alpha \,  \mathbf{\hat{n}} \times  \mathbf{\tilde{V}}  + (1-\cos \alpha) (\mathbf{\tilde{V}} \cdot \mathbf{\hat{n}}) \, \mathbf{\hat{n}} \,.
	\label{eq:a5}
	\end{equation} 
	%-------------------
	We can express the three sides of the triangle as 
	%-------------------
	\begin{equation}
	\ka=k_1 \, \Rabc \zh 
	\label{eq:a7}
	\end{equation}
	%-------------------
	%-------------------
	\begin{equation}
	\kb=t k_1( -\mu \Rabc \zh + \sqrt{1-\mu^2} \Rabc \xh)   
	\label{eq:a8}
	\end{equation}
	%-------------------
	and 
	%-------------------
	\begin{equation}
	\kc=- \ka - \kb 
	\label{eq:a9}
	\end{equation}
	%-------------------
	The  range $0 \le \alpha \le \pi$, $0 \le \beta \le \pi$ and $0 \le \gamma \le 2 \pi$  covers all rigid body rotations which also corresponds to all possible orientations of the triangle. We also need to integrate over all possible orientations of the triangle. The rotation group $SO(3)$ corresponds to the upper hemisphere of $S^3$ where we have the volume element 
	%-------------------
	\begin{equation}
	d^3 \R_{\alpha\beta\gamma}= \frac{1}{2} \sin^2 \frac{\alpha}{2}  \, \sin \beta \, d \alpha \, d \beta \, d \gamma \,.
	\label{eq:a10}
	\end{equation}
	%-------------------
	Integrating over all possible triangle orientations denoted by $\O$, we have
	%-------------------
	\begin{equation}
	\int_{\O}  d^3 \Rabc = \pi^2
	\label{eq:a11}
	\end{equation}
	%-------------------
	
	We see that six parameters $\alpha,\beta,\gamma,k_1,\mu,t$ are required to completely specify a triangle $\ka,\kb,\kc$. Alternatively we could use the three components of $\ka$ and $\kb$, six in total, to completely specify the same triangle. The  new  parameterization of the redshift space bispectrum introduced here, however, has the advantage that it allows us to fix the shape and size of the triangle by fixing the values of $k_1,t$ and $\mu$. In our parameterization it is possible to vary the orientation of the triangle while keeping the shape and size of the triangle fixed. 
	
	%***********************************************************************************
	\section{Quantifying Redshift Space Distortion}
	\label{sec:RSD}
	As mentioned earlier, the real space bispectrum does not depend on the triangle orientation, and we can express this as $B^r(k_1,\mu,t)$ which depends only on the size and shape of the triangle. Redshift space distortion introduces an addition feature where the redshift space bispectrum $B^s(\alpha,\beta,\gamma,k_1,\mu,t)$ also depends on the triangle orientation. Assuming the plane-parallel approximation along the line of sight (LoS) direction $\zh$, this dependence is through $\mu_1^2, \mu_2^2$ and $\mu_3^2$ where $\mu_a=\zh \cdot \mathbf{k}_a/k_a$ is the cosine of the angle between $\zh$ and the wave vector $\mathbf{k}_a$ with $a=1,2,3$ . The issue here is how to quantify the orientation dependence or anisotropy of  the redshift space bispectrum  which depends on $\mu_1,\mu_2,\mu_3$ {\it i.e.} the orientation of the three vectors $\ka,\kb,\kc$ with respect to the LoS direction $\zh$. 
	
	To place the issue in perspective we briefly recollect the redshift space power spectrum $P^s(\mu_1,k_1)$ where the anisotropy depends only on $\mu_1=\zh \cdot \mathbf{k}_1/k_1$.   In this case it is possible to quantify the anisotropy by using the multipole moments 
	%-------------------
	\begin{equation}
	P^s_{\ell}(k_1)=\frac{\int_{-1}^{1} \P_{\ell}(\mu_1)  P^s(\mu_1,k_1) \, d \mu_1}{\int_{-1}^{1} [\P_{\ell}(\mu_1)]^2  \, d \mu_1}
	\label{eq:b0}
	\end{equation}
	%-------------------
	where $\P_{\ell}(\mu_1)$ are the Legendre polynomials. The difficulty with the bispectrum is that the anisotropy depends on $\mu_1,\mu_2$ and  $\mu_3$ which refer to  the orientation of the three different vectors $\ka,\kb,\kc$ with respect to the LoS direction $\zh$. Further, $\mu_1,\mu_2$ and  $\mu_3$ are not independent but are related by the fact that they refer to particular triangles whose shape and  size are fixed, and whose orientation varies through different rigid body rotations  $\Rabc$.  
	
	Considering $\mu_1,\mu_2$ and  $\mu_3$, 
	using Equations~(\ref{eq:a7}),(\ref{eq:a8}) and (\ref{eq:a9}) we have 
	%-------------------
	\begin{equation}
	\mu_1=p_z
	\label{eq:b1}
	\end{equation}
	%-------------------
	%-------------------
	\begin{equation}
	\mu_2=-\mu p_z + \sqrt{1-\mu^2} p_x
	\label{eq:b2}
	\end{equation}
	%-------------------
	%-------------------
	\begin{equation}
	\mu_3=\frac{-[(1-t \mu) p_z + t \sqrt{1-\mu^2} p_x]}{\sqrt{1-2 t \mu +t^2}}
	\label{eq:b3}
	\end{equation}
	%-------------------
	where using \eqn~(\ref{eq:a5}) we have 
	%-------------------
	\begin{equation}
	p_z=\zh \cdot \Rabc \zh=\cos \alpha + (1-\cos \alpha) \,  n_z^2
	\label{eq:b4}
	\end{equation}
	%-------------------
	and 
	%-------------------
	\begin{equation}
	p_x=\zh \cdot \Rabc \xh = \sin \alpha \, n_y + (1-\cos \alpha ) \, n_x n_z \,.
	\label{eq:b5}
	\end{equation}
	%-------------------
	We can treat $p_z$ and $p_x$ as the Cartesian components of an unit vector $\ph=\Rabc^{-1} \zh$ with $p_y=\sqrt{1-p_x^2-p_z^2}$. Since $\mu_1,\mu_2$ and $\mu_3$ can all be expressed in terms of $\ph$, we use $B^s(\ph,k_1,\mu,t)$ to denote the orientation dependence of the redshift space bispectrum. Note that the entire $\alpha,\beta,\gamma$ dependence is contained within $\ph$ (Equations~\ref{eq:b4} and \ref{eq:b5}) which rotates as the orientation of the triangle is changed. We use spherical harmonics $Y^m_{\ell}(\ph)$ to quantify the orientation dependence or anisotropy of $B^s(\ph,k_1,\mu,t)$.

	We define the multipole moments  of the redshift space bispectrum  $B^s(\ph,k_1,\mu,t)$ as 
	%-------------------
	\begin{equation}
	\bar{B}^{m}_{\ell}(k_1,\mu,t)= \sqrt{\frac{(2 \ell+1)}{4\pi}}  \frac{\int_{\O} [Y^m_{\ell}(\ph)]^{*} B^s(\ph,k_1,\mu,t) \, d^3 \Rabc } {\int_{\O} \mid Y^m_{\ell}(\ph) \mid^2  \, d^3 \Rabc } \,.
	\label{eq:b7}
	\end{equation}
	%-------------------
	The normalization here has been chosen such that for $m=0$ we have  
	%-------------------
	\begin{equation}
	\bar{B}^{0}_{\ell}(k_1,\mu,t) = \frac{\int_{\O} \P_{\ell}(\mu_1) B^s(\ph,k_1,\mu,t)  \, d^3 \Rabc }{\int_{\O} [\P_{\ell}(\mu_1)]^2   \, d^3 \Rabc}
	\,.
	\label{eq:b8} 
	\end{equation}
	%-------------------
	which are exactly analogous to the multipole moments of the power spectrum defined in \eqn~(\ref{eq:b0}).  
	In the absence of redshift space distortion the monopole matches the real space bispectrum $(\bar{B}^0_0(k_1,\mu,t)=B^r(k_1,\mu,t))$,
	and all the higher multipole moments are zero ($\bar{B}^m_{\ell}(k_1,\mu,t)=0$ for $\ell >0$). 
	
	Note that the  integration here is over all possible orientations of the triangle. From the observational point of view, we need to identify the set $\T$ of all triangles $(\ka,\kb,\kc)$ which correspond to a fixed shape and size $k_1,\mu,t$, these essentially sample all possible orientations of the triangle. For each triangle, we  determine  $\kah=\ka/k_1=\Rabc \zh$ the unit vector along $\ka$ and $\kahp=\Rabc \xh$ the unit vector perpendicular to $\ka$ in the plane of the triangle $(\ka,\kb,\kc)$. We then have $p_z=\zh \cdot \kah$ and $p_x=\zh \cdot \kahp$, and we can use 
	%-------------------
	\begin{equation}
	\bar{B}^{m}_{\ell}(k_1,\mu,t)= \sqrt{\frac{ (2 \ell+1)}{4 \pi}} \frac{ \sum_{\T} [Y^m_{\ell}(\ph)]^{*} B^s(\ph,k_1,\mu,t)}{\sum_{\T} \mid Y^m_{\ell}(\ph) \mid^2}
	\label{eq:b8}
	\end{equation}
	%-------------------
	to estimate the various multipole moments. The present paper focuses on theoretical predictions and we do not apply the analysis to observational data here. We plan to address this in future work.

	Considering the theoretical perspective, we note that $\ph=\Rabc^{-1} \zh$ uniformly samples all possible directions if we consider the set of rigid body rotations $\O$. We can theoretically estimate the multipole moments using 
	%-------------------
	\begin{equation}
	\bar{B}^{m}_{\ell}(k_1,\mu,t) =\sqrt{\frac{(2 \ell +1)}{4 \pi}} \int [Y^m_{\ell}(\ph)]^* B^s(\ph,k_1,\mu) \, d\Omega_{\ph} 
	\label{eq:b9}
	\end{equation}
	%-------------------
	where  the $d\Omega_{\ph}$ integral is over  $4 \pi$ steradians.  In other words, instead over integrating over various orientations of the triangle $(\ka,\kb,\kc)$ we can theoretically predict the various multipole moments by considering a fixed triangle $(\kta,\ktb,\ktc)$ in the $x-z$ plane and integrating over all possible orientations of  $\ph=\Rabc^{-1} \zh$. In order to check that Equations~(\ref{eq:b7}) and (\ref{eq:b9}) give the same results,  
	we have explicitly  evaluated  several of the multipole moments $\bar{B}^m_{\ell}(k_1,\mu,t)$ using both these equations for the linear theory  considered in the next section.

	Like the power spectrum, here too the odd multipoles $(\ell=1,3,5,...)$  are all zero because the anisotropy only involves even powers of  $\mu_1,\mu_2$ and  $\mu_3$ in the plane parallel approximation. Further, since $B^s(\ph,k_1,\mu,t)$ is a real valued function with no explicit $p_y$ dependence, we have  
	$ \bar{B}^{-m}_{\ell}(k_1,\mu,t) = (-1)^m \bar{B}^{m}_{\ell}(k_1,\mu,t)$ where $\bar{B}^{m}_{\ell}(k_1,\mu,t)$ are all real. Considering 
	linear triangles $(\mu=1)$,  the vectors $\ka,\kb$ and $\kc$ are aligned in the same direction and we have $\mu_1=p_z=-\mu_2=-\mu_3$. In this case the redshift space bispectrum $B^s(\ph,k_1,\mu,t)$ has no 
	$p_x$ dependence, the anisotropy is completely quantified by the $m=0$ multipole moment and  the multipole moments with $m \neq 0$ are all zero. We however need both the $m=0$ and $m \neq 0$ multipole moments to completely quantify the anisotropy of $B^s(\ph,k_1,\mu,t)$ when the three vectors $\ka,\kb$ and $\kc$ are not aligned  {\it i.e.} $(\mu <1)$. 
	
	%***********************************************************************************
	\section{Linear Redshift Space Distortion}
	\label{sec:linear-RSD}
	Here we consider the linear theory of redshift space distortion \citep{hamilton-98-rsd-review} where the fluctuations in redshift space $\Delta^s(\ka)$
	is related to the corresponding fluctuation $\Delta^r(\ka)$ in real space  as 
	%-------------------
	\begin{equation}
	\Delta^s(\ka)=(1 + \beta_1 \mu_1^2) \Delta^r(\ka)
	\label{eq:c1}
	\end{equation}
	%-------------------
	with $\beta_1=f(\Omega_m)/b_1$ being the linear redshift distortion parameter. Here $f(\Omega_m)$ is the linear growth rate of density perturbations and $b_1$ is the linear bias parameter. 
	We then have 
	%-------------------
	\begin{equation}
	B^s(\ph,k_1,\mu,t)= (1+ \beta_1 \mu_1^2) (1+ \beta_1 \mu_2^2) (1+ \beta_1 \mu_3^2)  B^r(k_1,\mu,t) \,.
	\label{eq:c2}
	\end{equation}
	%-------------------
	Considering sufficiently large length scales which are in the  linear regime, \eqn~(\ref{eq:c2}) is a valid model for the bispectrum arising from primordial non-Gaussianity \citep{bartolo04-PNG-rev}. We can define the enhancement factor 
	%-------------------
	\begin{equation}
	\RB^m_{\ell}(\beta_1,\mu,t)=\frac{\bar{B}^m_{\ell}(k_1,\mu,t)}{B^r(k_1,\mu,t)}
	\label{eq:c3}
	\end{equation}
	%-------------------
	which only depends on $\beta_1$ and the shape of the triangle.
	We can calculate this using 
	%-------------------
	\begin{eqnarray}
	\RB^m_{\ell}(\beta_1,\mu,t)&=& \sqrt{\frac{(2 \ell +1)}{4 \pi}} 
	\int  d\Omega_{\ph}  \, [Y^m_{\ell}(\ph)]^* \nn \times \\ && \, (1+ \beta_1 \mu_1^2) (1+ \beta_1 \mu_2^2) (1+ \beta_1 \mu_3^2) \,.
	\label{eq:c4}
	\end{eqnarray}
	%-------------------
	As mentioned earlier, only the even multipole moments are non-zero. In this case only the first four even terms $\ell=0,2,4,6$ are non-zero. 
	
	%-------------------
	\begin{figure}
		\centering
		\includegraphics[width=0.47\textwidth,angle=0]{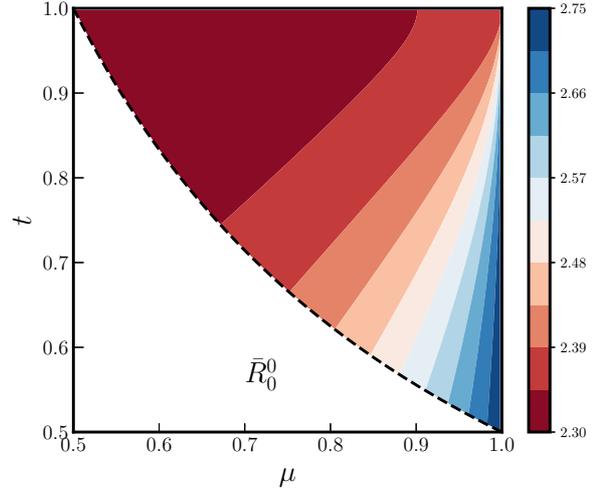}
		\caption{This shows the monopole enhancement factor $\RB^0_0(\mu,t)$ for triangles of different shapes.}
		\label{fig:R00}
	\end{figure}
	%-------------------
	
\begin{figure}
		\centering
		\includegraphics[width=0.45\textwidth,angle=0]{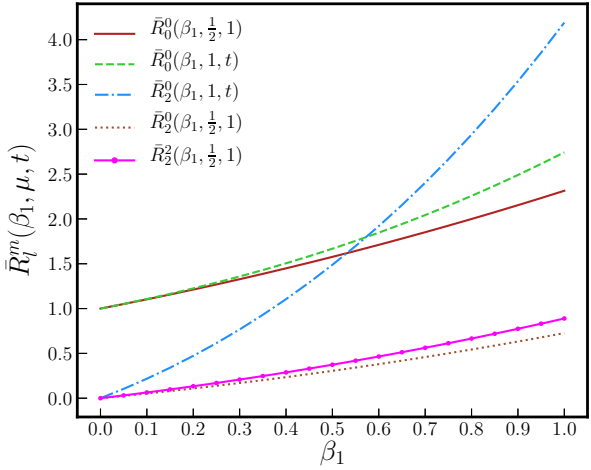}
		\caption{This shows $\beta_1$ dependence of $\RB^0_0(\beta_1,\mu,t)$, $\RB^0_2(\beta_1,\mu,t)$   and $\RB^2_2(\beta_1,\mu,t)$ for the specific triangle configurations indicated in the figure.}
		\label{fig:Rlm_beta}
	\end{figure}

	Considering the monopole first, we have 
	%-------------------
	\begin{eqnarray}
	\RB^0_0(\beta_1,\mu,t) &=& 1+\beta_1+\frac{3 \beta_1^2}{5}+\frac{\beta_1^3}{7}-\nn\\ &&  \frac{4 \beta_1^2 (3 \beta_1+7) \left(1-\mu ^2\right) \left(t^2-\mu  t+1\right)}{105 \left(t^2-2 \mu  t+1\right)} \,.
	\label{eq:c5}
	\end{eqnarray}
	%-------------------
	Figure~\ref{fig:R00} shows $\RB^0_0(\beta_1,\mu,t)$ for different triangle shapes considering $\beta_1=1$.  We see that redshift space distortion enhances the bispectrum monopole $(\RB^0_0(\beta_1,\mu,t)>1)$ for  triangles of all shapes. This enhancement is maximum  for linear triangles.  This corresponds to the first term in the $r.h.s.$ of \eqn~(\ref{eq:c5}) which is independent  of $t$, the second term having value zero when $\mu=1$.  The value of the enhancement factor  $\RB^0_0(\beta_1,\mu,t)$ falls off  away from the  $\mu=1$ line {\it i.e.} the three vectors $\ka,\kb,\kc$ are no longer aligned, and it is  minimum for $\mu=0.5$ which corresponds to equilateral triangles. However, the difference between the maximum value ($96/35$) and the minimum value ($81/35$) is not very large $((96-81)/81\approx 18 \%)$ for $\beta_1=1$, and it is smaller  for other values $\beta_1<1$. Figure~\ref{fig:Rlm_beta} shows $\RB^0_0(\beta_1,\mu,t)$ as a function of $\beta_1$ for the two extreme triangle shapes $\mu=1$ and $\mu=0.5$ respectively.
	
	%-------------------
	%-------------------
	
	%-------------------
	\begin{figure*}
		\centering
		\includegraphics[width=1\textwidth,angle=0]{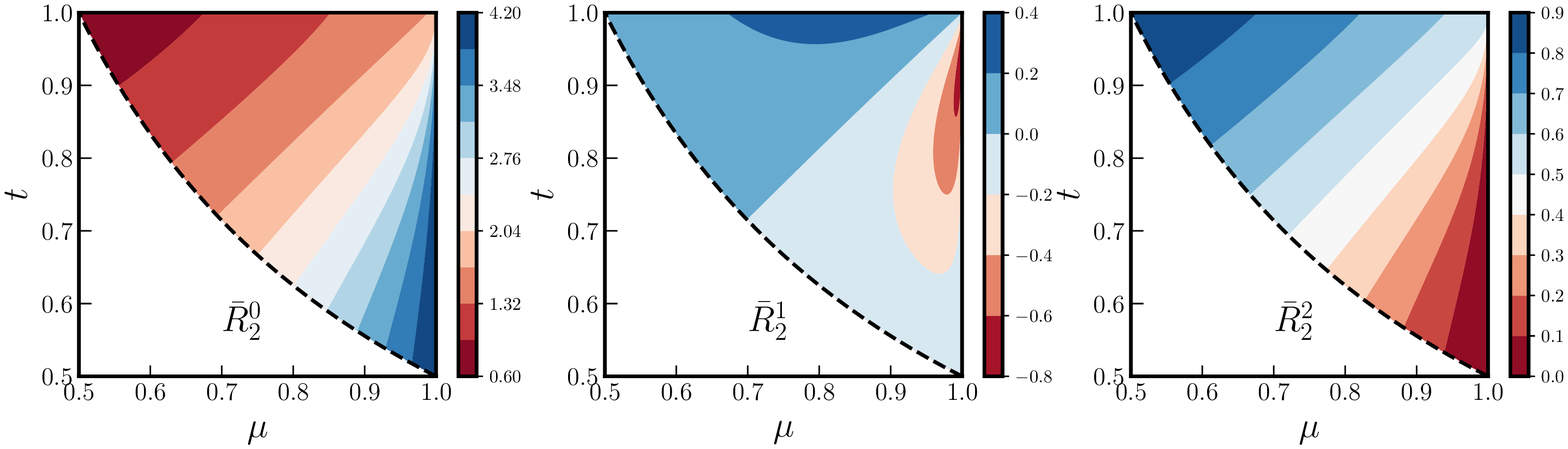}
		\caption{This shows the three quadrupole moments $\RB^0_2(\beta_1,\mu,t)$, $\RB^1_2(\beta_1,\mu,t)$ and  $\RB^2_2(\beta_1,\mu,t)$ as functions of $(\mu,t)$ for $\beta_1=1$.}
		\label{fig:R2m}
	\end{figure*}
	%-------------------

	We now consider the different $\ell >0$  multipole moments which characterise the anisotropy of the redshift space bispectrum. From \eqn~(\ref{eq:c2}) we see that 
	in linear theory all of these  are expected to be polynomials of the form $a \beta_1 + b \beta_1^2 + c  \beta_1^3$  where $a,b,c$   depend  on $\ell,m$ and the shape of the triangle $(\mu,t)$. The multipole moments with $m=0$  are only sensitive to $p_z=\mu_1$. These multipole moments with $m=0$ are all expected to have the maximum value for linear triangles $(\mu=1)$  where  $\ka,\kb,\kc$  are aligned and $\mu_1^2=\mu_2^2=\mu_3^2=p_z^2$  (Equations~\ref{eq:b1}, \ref{eq:b2} and \ref{eq:b3}),  and they are expected to have the minimum value for equilateral triangles $(\mu=0.5)$.  
	The multipole moments with $m \neq 0$ quantify the joint $p_z,p_x$ dependence of $B^s(\ph,k_1,\mu,t)$. The  linear triangles  have no $p_x$ dependence, and in this case    the multipole moments with $m \neq 0$ are all predicted to be zero.  Further, 
	the highest power of $p_x$ in \eqn~(\ref{eq:c2}) is $p_x^4$, and this implies that the multipole moments with $m>4$ are all zero.

	The  quadruple $\RB^m_2(\beta_1,\mu,t)$ has  three independent terms $m=0,1$ and $2$ for which 
	the results are presented below. 
	%-------------------
	\begin{eqnarray}
	\RB^0_2(\beta_1,\mu,t) &=& \frac{2}{21} \beta_1 \left(5 \beta_1^2+18 \beta_1+21\right) -
	\frac{\beta_1 \left(1-\mu ^2\right)}{21 \left(t^2-2 \mu  t+1\right)}\nn\\
	&&\times\Big[9 \beta_1^2+22 \beta_1+2 t^2 (\beta_1^2\left(2 \mu ^2+5\right)+20 \beta_1\nn\\ 
	&&+21)-2 \left(7 \beta_1^2+20 \beta_1+21\right) \mu  t+21\Big]\, ,
	\label{eq:c6}
	\end{eqnarray}
	%======================
	\begin{eqnarray}
	\RB^1_2(\beta_1,\mu,t) &=& \frac{1}{21 \left(t^2-2 \mu  t+1\right)}\sqrt{\frac{2}{3}} \beta_1 \sqrt{1-\mu ^2} (2 \mu  t-1)\nn\\
	&&\times \Big[t \left(\beta_1^2 \left(2 \mu ^2+3\right)+18 \beta_1+21\right)\nn\\
	&& -\left(5 \beta_1^2+18 \beta_1+21\right) \mu \Big]\, ,
	\label{eq:c7}
	\end{eqnarray}
	%======================
	\begin{eqnarray}
	\RB^2_2(\beta_1,\mu,t) &= & \frac{1}{21 \sqrt{6} \left(t^2-2 \mu  t+1\right)}\beta_1 \left(1-\mu ^2\right)\nn\\
	&& \times\Big[\beta_1 (\beta_1+6)+2 t^2 (\beta_1 \left(2 \beta_1 \mu ^2+\beta_1+12\right)\nn\\
	&&+21)-6 (\beta_1 (\beta_1+4)+7) \mu  t+21\Big]\, .
	\label{eq:c8}
	\end{eqnarray}
	%-------------------
	
	Considering $\RB^0_2(\beta_1,\mu,t)$  (\eqn~(\ref{eq:c6})) and left panel of Figure~\ref{fig:R2m}), we see that, like the monopole, and  as expected,  this  is maximum for the linear triangle $(\mu=1)$ independent of $t$. This corresponds to the first term in the $r.h.s.$ of \eqn~(\ref{eq:c6}) which is independent  of $t$, the second term having value zero when $\mu=1$. The value of   $\RB^0_2(\beta_1,\mu,t)$ falls off away from the line corresponding to  $\mu$=1, and it is  minimum for $\mu=0.5$ which corresponds to equilateral triangles. 
	Unlike the monopole, we find a rather large difference between the maximum value ($88/21$) and the minimum value ($61/84$) of  the  quadrupole component  $\RB^0_2(\beta_1,\mu,t)$ for $\beta_1=1$. Figure~\ref{fig:Rlm_beta}  shows the $\beta_1$ dependence of $\RB^0_2(\beta_1,\mu,t)$ for the two extreme triangle shapes $\mu=1$ and $\mu=0.5$ respectively. 
	
	Considering the two other quadrupole components $\RB^1_2(\beta_1,\mu,t)$  and $\RB^2_2(\beta_1,\mu,t)$ respectively presented in Equations~(\ref{eq:c7}) and (\ref{eq:c8}) and shown in the center and right panels of Figure~\ref{fig:R2m},  we see that unlike $\RB^0_0(\beta_1,\mu,t)$ and $\RB^0_2(\beta_1,\mu,t)$, these are zero for linear triangles $(\mu=1)$ independent of the value of $t$. Further,  $\RB^1_2(\beta_1,\mu,t)$  is also zero for the S isosceles triangles $(2 t \mu=1)$  and the equilateral triangle $(\mu=0.5)$. We see that $\RB^1_2(\beta_1,\mu,t)$  has negative values in the  lower right  region $\mu > t$ corresponding to obtuse triangles. We have  $\RB^1_2(\beta_1,\mu,t) \approx 0$ around the line $\mu=t$ which corresponds to right-angle triangles. However the exact curve 
	%-------------------
	\begin{equation}
	t=\frac{5 \beta_1^2 + 18 \beta_1 + 21}{(2 \mu^2 +3) \beta_1^2 + 18 \beta_1 +21} 
	\label{eq:c7a}
	\end{equation}
	%-------------------
	along which $\RB^1_2(\beta_1,\mu,t)=0$ is slightly above $\mu=t$ and it also encompasses  some of the acute triangles. $\RB^1_2(\beta_1,\mu,t)$ has positive values for  most of upper left  region 
	$\mu <t$ which corresponds to  acute triangles. We also note that the difference between the maximum and the minimum values of $\RB^1_2(\beta_1,\mu,t)$  is of the order of unity for $\beta_1=1$, this will be less for lower values of $\beta_1$.

	Considering $\RB^2_2(\beta_1,\mu,t)$, we see that it has values in the range $0 \le \RB^2_2(\beta_1,\mu,t)$, and it has a  maximum value of $61/(28 \sqrt{6})$ for equilateral triangles. Figure~\ref{fig:Rlm_beta}  shows the $\beta_1$ dependence of $\RB^2_2(\beta_1,\mu,t)$ for equilateral triangles.  We note that both $\RB^0_2(\beta_1,\mu,t)$ and $\RB^2_2(\beta_1,\mu,t)$  show very similar $\beta_1$ dependence for equilateral triangles. Also, the contour plots corresponding to $\RB^0_2(\beta_1,\mu,t)$ and $\RB^2_2(\beta_1,\mu,t)$  (the left and right panels of Figure~\ref{fig:R2m} respectively) show very similar patterns, though the values are quite different.

\begin{figure*}
		\centering
		\includegraphics[width=1\textwidth,angle=0]{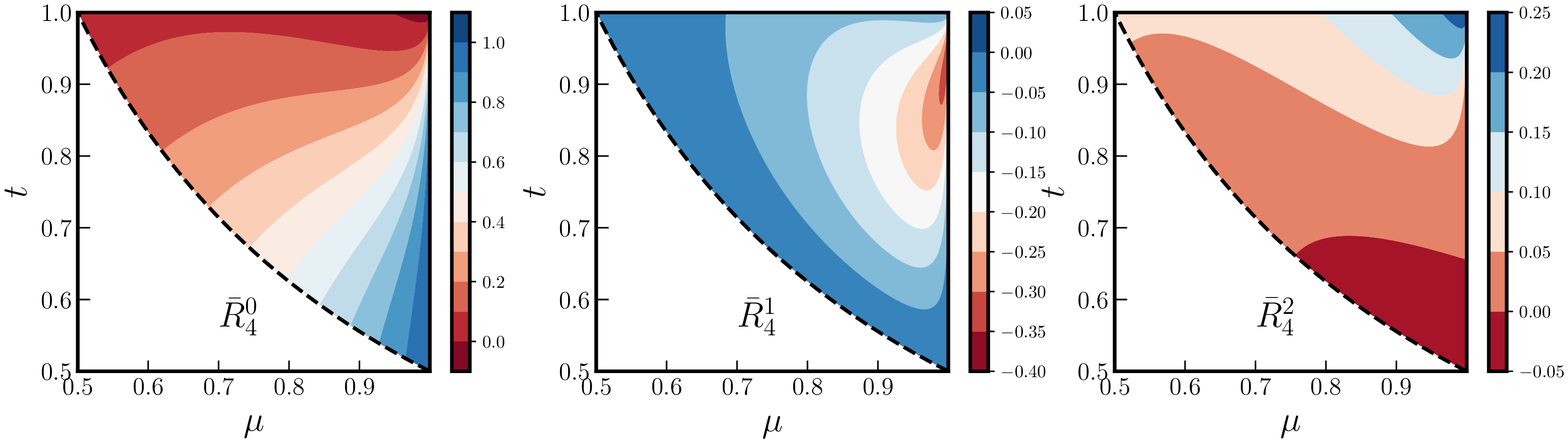}
		\caption{This shows the three quadrupole moments $\RB^0_4(\beta_1,\mu,t)$, $\RB^1_4(\beta_1,\mu,t)$ and  $\RB^2_4(\beta_1,\mu,t)$ as functions of $(\mu,t)$ for $\beta_1=1$.}
		\label{fig:R4m}
	\end{figure*}

	We next consider the hexadecapole $\RB^m_4(\beta_1,\mu,t)$ for which we have five independent terms $m=0,1,2,3,4$.  We only discuss  the  first three hexadecapole moments  $(m=0,1,2)$  
	for which the results are shown in Figure~\ref{fig:R4m} for $\beta_1=1$. 
	The other haxadecapole moments $(m=3,4)$ have relatively small values  $\mid \RB^m_4(\beta_1,\mu,t)\mid <0.1$,  and for completeness we have presented these in Appendix~\ref{appendix1}. Considering  $\RB^0_4(\beta_1,\mu,t)$ (\eqn~(\ref{eq:c9}) and left panel of Figure~\ref{fig:R4m}), as expected, we see that  this has the maximum value ($384/385$ for $\beta_1=1$) for linear triangles. This corresponds to the first term in the $r.h.s.$ of  \eqn~(\ref{eq:c9}), and the second term in the $r.h.s$  is zero for $\mu=1$. The values of $\RB^0_4(\beta_1,\mu,t)$ falls off away from the line corresponding to $\mu=1$, and it is minimum for equilateral triangles where it has a value $81/1540$ for $\beta_1=1$. 
	%-------------------
	\begin{eqnarray}
	\RB^0_4(\beta_1,\mu,t) &=&\frac{24}{385} \beta_1^2 (5 \beta_1+11)- \frac{\beta_1^2 \left(1-\mu ^2\right)}{385 \left(t^2-2 \mu  t+1\right)}\times\nn\\
	&&\Big[8 (17 \beta_1+33)+t^2 (3 \beta_1 \left(65 \mu ^2+47\right)\nn\\
	&& +385 \mu ^2+319)-16 (21 \beta_1+44) \mu  t\Big]
	\label{eq:c9}
	\end{eqnarray}
	%-------------------
	
	We next consider the multipole moments $\RB^1_4(\beta_1,\mu,t)$ and $\RB^2_4(\beta_1,\mu,t)$ for which the results are presented in Equations~(\ref{eq:c10}) and (\ref{eq:c11}) and  shown in the center and right panels of Figure~\ref{fig:R4m} respectively.  As expected, both of these are zero for linear triangles $(\mu=1)$. Further,  $\RB^1_4(\beta_1,\mu,t)$  is also zero for the S isosceles triangles and  equilateral triangles. This multipole moment has predominantly negative values with the minimum value occurring near the stretched triangle configuration, and the difference  between  the maximum and minimum values is around $0.5$ for $\beta_1=1$. Considering $\RB^2_4(\beta_1,\mu,t)$, we see that this is predominantly positive 
	with maximum value near the squeezed triangle configuration and   small negative values near the stretched triangle configuration. 
	The difference between  the maximum and minimum values gets smaller as we go to larger values of $m$, and considering   $\RB^2_4(\beta_1,\mu,t)$ it is around $0.25$ for $\beta_1=1$. 
	%-------------------
	\begin{eqnarray}
	\RB^1_4(\beta_1,\mu,t) &= &\frac{1}{77 \sqrt{5} \left(t^2-2 \mu  t+1\right)}\beta_1^2 \sqrt{1-\mu ^2} (2 \mu  t-1)\nn\\ 
	&&\Big[t \left(\beta_1 \left(37 \mu ^2+3\right)+77 \mu ^2+11\right)\nn\\
	&& -8 (5 \beta_1+11) \mu \Big]
	\label{eq:c10}
	\end{eqnarray}
	%-------------------
	%-------------------
	\begin{eqnarray}
	\RB^2_4(\beta_1,\mu,t) &=& \frac{1}{77 \left(t^2-2 \mu  t+1\right)}\sqrt{\frac{2}{5}} \beta_1^2 \left(1-\mu ^2\right) \Big[6 \beta_1+\nn\\
	&& t^2 \left(\beta_1 \left(31 \mu ^2+5\right)+77 \mu ^2+11\right)-\nn\\
	&& 4 (9 \beta_1+22) \mu  t+22\Big]
	\label{eq:c11}
	\end{eqnarray}
	%-------------------
	%-------------------
	%-------------------
	%-------------------
	\begin{figure}
		\centering
		\includegraphics[width=0.45\textwidth,angle=0]{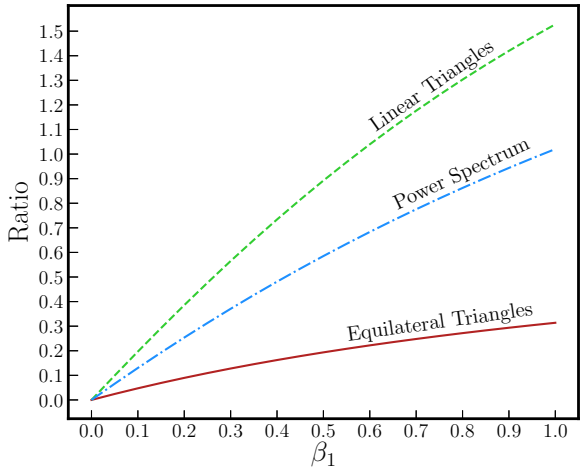}
		\caption{This shows the quadrupole ($\bar{B}^0_{2}$) to monopole ($\bar{B}^0_{0}$) ratio as a function of $\beta_1$ for both linear as well as equilateral triangles, in addition to the quadrupole ($P^s_2$) to monopole ($P^s_0$) ratio for the redshift space power spectrum.}
		\label{fig:quad_by_mono_beta}
	\end{figure}
	%-------------------
	The multipole moments with $\ell=4$ and $m=3,4,5$, and all the multipole moments with $\ell=6$  have small values $\mid \RB^m_{\ell}(\beta_1,\mu,t) \mid <0.1$ and we have shown these in the Appendix for completeness. 
	
	%***********************************************************************************
	\section{Discussion}
	\label{sec:discussion}
	Here we have proposed a method to quantify the redshift space bispectrum by decomposing it into  multipole moments  $\bar{B}^m_{\ell}(k_1,\mu,t)$. In this work we have carried out a detailed analysis of the situation where the bispectrum is produced by primordial non-Gaussianity, and is assumed to evolve according to linear theory. We provide explicit analytical expressions relating   all the non-zero multipole moments to the real space bispectrum $B^r(k_1,\mu,t)$. Considering the monopole (\eqn~(\ref{eq:c5}))
	we find that this is enhanced with respect to $B^r(k_1,\mu,t)$, the enhancement factor depending on the shape of the triangle and the value of $\beta_1$. The enhancement factor is maximum $(1+\beta_1 +(3/5)  \beta_1^2 +(1/7)\beta_1 ^3)$ for linear triangles and minimum $(1+\beta_1 +(3/10)  \beta_1^2 +(1/70)\beta_1 ^3)$ for equilateral triangles, and has values in between these for triangles of other shapes.   
	
	 The bisprectrum measured from redshift surveys will, in general, be a combination of  the bispectrum of  primordial non-Gaussianity (PNG) and the bispectrum  induced by the non-linear gravitational evolution (NG). The PNG parameter $f_{\rm NL}$  is tightly bound  by the bispectrum of the CMB temperature and polarization anisotropies \citep{planck18png},  and one expects the measured bispectrum to be dominated by the NG contribution at the  length-scales of the  Baryon Acoustic Oscillation (BAO) and smaller.  The analysis presented in this paper, which is the first in a series of papers, is restricted to the PNG bispectrum. The  same formalism can be used to quantify the redshift space distortion of the NG bispectrum which will be presented in  a subsequent paper. It may be noted that the relative contribution from the PNG bispectrum becomes significant at high redshifts and  at $k \le k_{eq}$    where  $k_{eq}$ is the comoving wave-number corresponding to  epoch of matter radiation equality. 
	The enhancement due to $\RB^0_{0}(\beta_1,\mu,t)$  (Figure~\ref{fig:Rlm_beta}) increases the prospects of detecting the PNG bispectrum  using redshift surveys. Considering a future scenario where 	we have a measurement of the PNG bispectrum 
	$\bar{B}^{m}_{\ell}(k_1,\mu,t)$ 	from redshift surveys, 	it is necessary to account for the fact that this will yield an estimate of $\RB^0_{0}(\beta_1,\mu,t) \times f_{\rm NL}$ which depends on $\beta_1$ and the shape of the triangle. 
	It will be important to account  for this  enhancement in order to infer   a precise  $f_{\rm NL}$ value.

	The quadrupole and the other higher multipole moments explicitly quantify the anisotropy introduced by redshift space distortion. The leading multipole here is the quadrupole component  $\bar{B}^0_{2}(k_1,\mu,t)$. Expressing this in terms of  $B^r(k_1,\mu,t)$, we find that this has a a maximum value of  $\frac{2}{21} \beta_1 \left(5 \beta_1^2+18 \beta_1+21\right)$ for linear triangles and a minimum value of $\frac{1}{84} \beta_1 \left(\beta_1^2+18 \beta_1+42\right)$ for equilateral triangles, and has values in between these for triangles of other shapes. We see that $\bar{B}^0_{2}(k_1,\mu,t)$ has a strong $\beta_1$ dependence (Figure~\ref{fig:Rlm_beta}), particularly for linear triangles. This holds the possibility of allowing us to use measurements of the redshift space bispectrum to estimate 
	$\beta_1$, analogous to the $\beta_1$  estimates from the redshift space power spectrum. To illustrate this we consider a situation where the measured bispectrum is dominated by the contribution from primordial non-Gaussianity, the induced bispectrum caused by the non-linear evolution making a subdominant contribution. Figure~\ref{fig:quad_by_mono_beta} shows the quadrupole ($\bar{B}^0_{2}$) to monopole ($\bar{B}^0_{0}$) ratio as a function of $\beta_1$ for both linear as well as equilateral triangles, in addition to the quadrupole to monopole ratio for the power spectrum. We see that this ratio is more sensitive to $\beta_1$ for  linear triangles  as compared to the power spectrum, whereas for equilateral triangles it is less sensitive than the power spectrum. This ratio is sensitive to both  $\beta_1$ and the  shape of the triangle, and it should be possible to improve the $\beta_1$ estimates by jointly modelling the power spectrum and the bispectrum \citep{gilmarin17-PS-BS-RSD} instead of using the power spectrum alone.  RSD

	The multipole moment $\RB^m_{\ell}(k_1,\mu,t)$ shows a very interesting shape dependence where  it is positive for acute triangles and negative for obtuse triangles. The various other multipole moments each  shows a  very distinctive shape dependence.  Several of these $(\ell,m)=(2,1),(2,2),(4,0))$ show variations of order unity  if the shape of the triangle is changed for $\beta_1=1$. A  couple more  $(\ell,m)=(4,1),(4,2)$ show 
	variations of order $\sim 0.4$. These characteristic  variations, if measured, will  impose further constraints on $\beta_1$. Further, such characteristic   shape dependencies may also prove to be a distinctive feature of linear primordial non-Gaussian fluctuations.

	The present paper is entirely restricted to a linear theory analysis of the bispectrum arising from primordial non-Gaussianity. We however expect the induced bispectrum arising from non-linear evolution to dominate,  particularly at low redshifts. We plan to address the theoretical predictions for this scenario in future work. 
	
	%******************************APPENDIX**********************************
	\appendix
	\section{Higher Multipoles}
	\label{appendix1}
	%-------------------
	\begin{figure*}
		\centering
		\includegraphics[width=0.66\textwidth,angle=0]{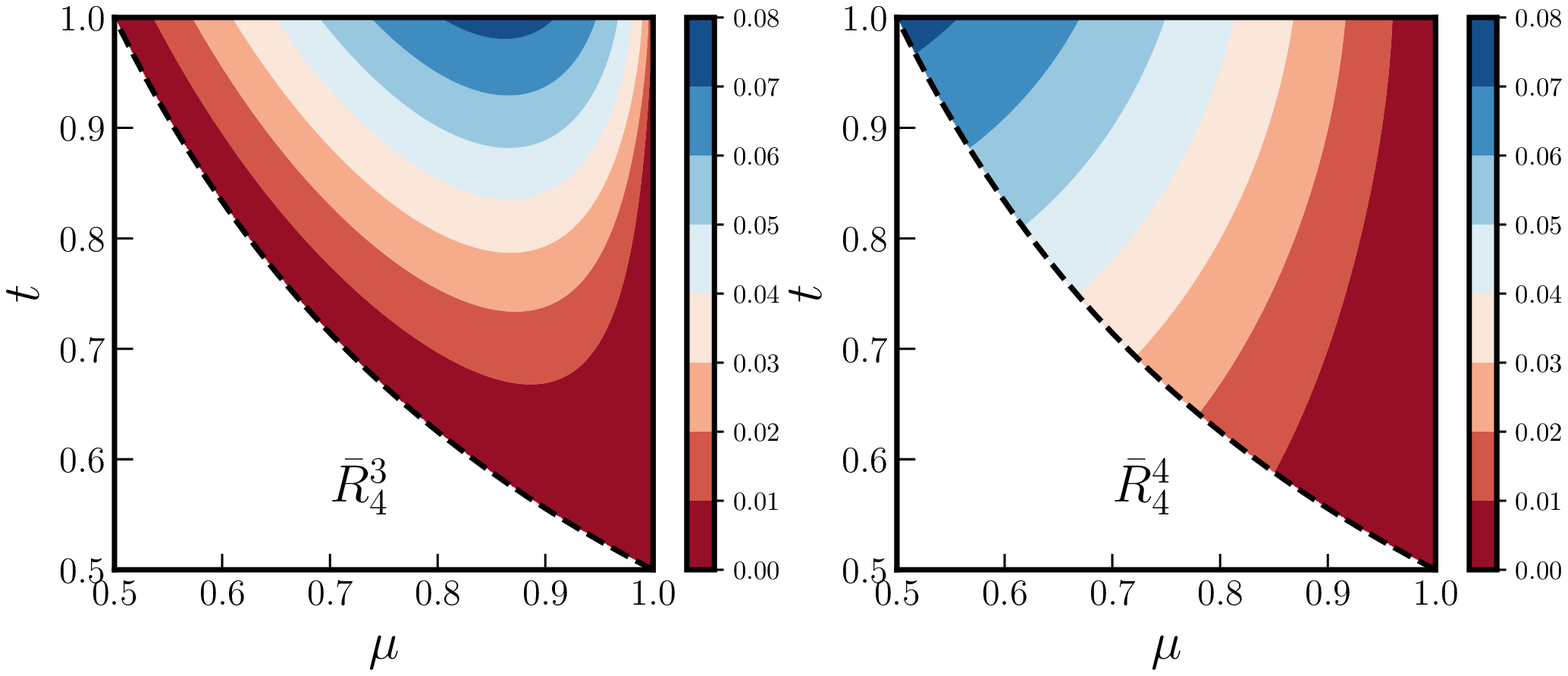}
		\caption{This shows the three quadrupole moments $\RB^3_4(\beta_1,\mu,t)$ and  $\RB^4_4(\beta_1,\mu,t)$ as functions of $(\mu,t)$ for $\beta_1=1$.}
		\label{fig:R4ma}
	\end{figure*}
	%-------------------
	
	Here we first consider two of the hexadecapole moments $\RB^3_4(\beta_1,\mu,t)$ and  $\RB^4_4(\beta_1,\mu,t)$ 
	for which the analytical expressions are respectively presented in Equations~(\ref{eq:aa1}) and (\ref{eq:aa2}), and the results shown for $\beta_1=1$ in Figure~\ref{fig:R4ma}. 
	%-------------------
	\begin{eqnarray}
	\RB^3_4(\beta_1,\mu,t) &=& \frac{\beta_1 ^2 (3 \beta_1 +11) \left(1-\mu ^2\right)^{3\over2} t (2 \mu  t-1)}{11 \sqrt{35} \left(t^2-2 \mu  t+1\right)}
	\label{eq:aa1}
	\end{eqnarray}
	%-------------------
	%-------------------
	\begin{eqnarray}
	\RB^4_4(\beta_1,\mu,t) &=&\frac{\beta_1^2 (\beta_1 +11) \left(1-\mu ^2\right)^2 t^2}{11 \sqrt{70} \left(t^2-2 \mu  t+1\right)}
	\label{eq:aa2}
	\end{eqnarray}
	%-------------------

	%-------------------
	\begin{figure*}
		\centering
		\includegraphics[width=1\textwidth,angle=0]{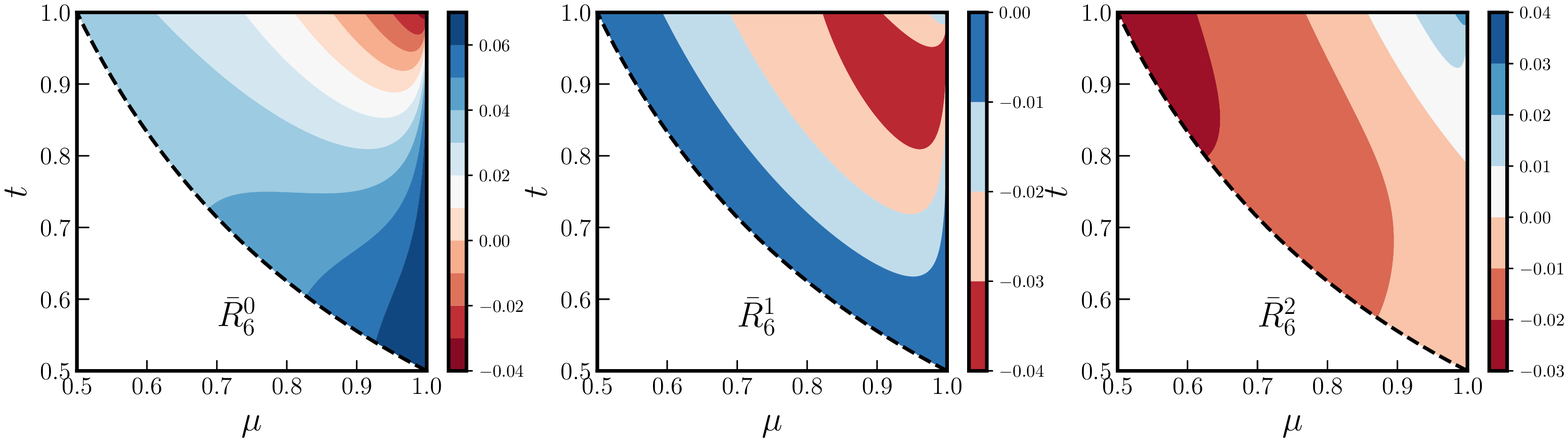}
		\caption{This shows the three multipole  moments $\RB^0_6(\beta_1,\mu,t)$, $\RB^1_6(\beta_1,\mu,t)$ and  $\RB^2_6(\beta_1,\mu,t)$ as functions of $(\mu,t)$ for $\beta_1=1$.}
		\label{fig:R6m}
	\end{figure*}
	%-------------------
	%-------------------
	\begin{figure*}
		\centering
		\includegraphics[width=0.7\textwidth,angle=0]{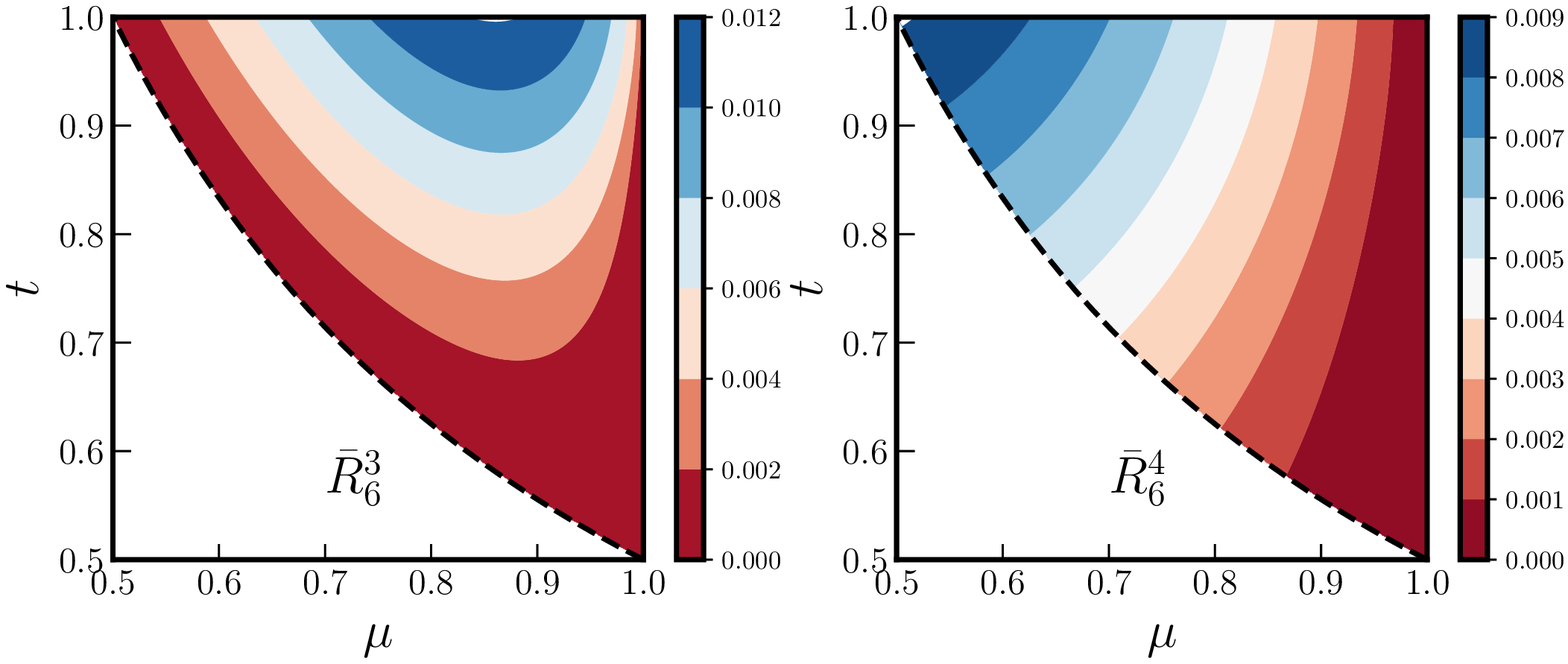}
		\caption{This shows the two multipole  moments $\RB^3_6(\beta_1,\mu,t)$ and  $\RB^4_6(\beta_1,\mu,t)$ as functions of $(\mu,t)$ for $\beta_1=1$.}
		\label{fig:R6ma}
	\end{figure*}
	%-------------------
	
	Considering  $\ell=6$  and  $m=0$ to $4$, the analytic expressions for the various multipole moments  are presented in Equations~(\ref{eq:aa3}) to   (\ref{eq:aa7}). For these the results  are shown in Figures  \ref{fig:R6m} and \ref{fig:R6ma}. As mentioned earlier, $R_6^5(\beta_1,\mu,t)$ and $R_6^6(\beta_1,\mu,t)$ are both  zero.
	%-------------------
	\begin{eqnarray}
	\RB^0_6(\beta_1,\mu,t)& =&\frac{16 \beta_1^3}{231}-\frac{2 \beta_1^3 \left(1-\mu ^2\right)}{231}\times\nn\\
	&&\frac{ \left(5 \left(7 \mu ^2+1\right) t^2-40 \mu  t+12\right)}{ \left(t^2-2 \mu  t+1\right)}
	\label{eq:aa3}
	\end{eqnarray}
	%-------------------
	%-------------------
	\begin{eqnarray}
	\RB^1_6 (\beta_1,\mu,t) &=& 2 \sqrt{\frac{2}{21}} \beta_1^3 \sqrt{1-\mu ^2} (2 \mu  t-1)\times \nn\\
	&&\frac{\left(\left(7 \mu ^2-3\right) t-4 \mu \right)}{33 \left(t^2-2 \mu  t+1\right)}
	\label{eq:aa4}
	\end{eqnarray}
	%-------------------
	%-------------------
	\begin{eqnarray}
	\RB^2_6 (\beta_1,\mu,t) = \frac{8 \beta_1^3 \left(1-\mu ^2\right) \left(\left(7 \mu ^2-1\right) t^2-6 \mu  t+1\right)}{33 \sqrt{105} \left(t^2-2 \mu  t+1\right)} 
	\label{eq:aa5}
	\end{eqnarray}
	%-------------------
	%-------------------
	\begin{eqnarray}
	\RB^3_6(\beta_1,\mu,t) = \frac{4 \beta_1^3 \left(1-\mu ^2\right)^{3/2} t (2 \mu  t-1)}{11 \sqrt{105} \left(t^2-2 \mu  t+1\right)}
	\label{eq:aa6}
	\end{eqnarray}
	%-------------------
	%-------------------
	\begin{eqnarray}
	\RB^4_6(\beta_1,\mu,t) = \sqrt{\frac{2}{7}}\frac{ \beta_1^3 \left(1-\mu ^2\right)^2 t^2}{33 \left(t^2-2 \mu  t+1\right)}
	\label{eq:aa7}
	\end{eqnarray}
	%-------------------

	%%%%%%%%%%%%%%%%%%%%%%%%%%%%%%%%%%%%%%%%%%%%%%%%%
	
	
	
	\bibliographystyle{mnras}
	\bibliography{reference} % if your bibtex file is called example.bib

	
	%%%%%%%%%%%%%%%%%%%%%%%%%%%%%%%%%%%%%%%%%%%%%%%%%%
	
	%%%%%%%%%%%%%%%%% APPENDICES %%%%%%%%%%%%%%%%%%%%%
	
	%%%%%%%%%%%%%%%%%%%%%%%%%%%%%%%%%%%%%%%%%%%%%%%%%%

	% Don't change these lines
	\bsp	% typesetting comment
	\label{lastpage}
\end{document}